\documentclass[%
 preprint,
superscriptaddress,
 amsmath,amssymb,
 aps,
prstab,
floatfix,
]{revtex4-2}

\usepackage{graphicx}
\usepackage{dcolumn}
\usepackage{bm}
\usepackage{siunitx}

\begin{document}


\title{A NEURAL NETWORK MODEL OF A QUASI-PERIODIC ELLIPTICALLY POLARIZING UNDULATOR IN UNIVERSAL MODE\\}

\author{Ryan Sheppard}
\affiliation{%
 Canadian Light Source, 44 Innovation Blvd., Saskatoon, Saskatchewan, Canada\\
}%
\affiliation{%
 University of Saskatchewan, Department of Physics and Engineering Physics, 116 Science Place, Saskatoon, Saskatchewan, Canada\\
}%
\affiliation{
McGill University, 817 Sherbrooke St. West, Montreal, Quebec, Canada
}
\author{Cameron Baribeau}
\affiliation{%
 Canadian Light Source, 44 Innovation Blvd., Saskatoon, Saskatchewan, Canada\\
}%
\author{Tor Pedersen}
\affiliation{%
 Canadian Light Source, 44 Innovation Blvd., Saskatoon, Saskatchewan, Canada\\
}%
\author{Mark Boland}
\affiliation{%
 Canadian Light Source, 44 Innovation Blvd., Saskatoon, Saskatchewan, Canada\\
}%
\affiliation{%
 University of Saskatchewan, Department of Physics and Engineering Physics, 116 Science Place, Saskatoon, Saskatchewan, Canada\\
}%
\author{Drew Bertwistle}%
\email{drew.bertwistle@lightsource.ca}
\affiliation{%
 Canadian Light Source, 44 Innovation Blvd., Saskatoon, Saskatchewan, Canada\\
}%
\affiliation{%
 University of Saskatchewan, Department of Physics and Engineering Physics, 116 Science Place, Saskatoon, Saskatchewan, Canada\\
}%

\date{\today}

\begin{abstract}
Machine learning has recently been applied and deployed at several light source facilities in the domain of Accelerator Physics. We introduce an approach based on machine learning to produce a fast-executing model that predicts the polarization and energy of the radiated light produced at an insertion device. This paper demonstrates how a machine learning model can be trained on simulated data and later calibrated to a smaller, limited measured data set, a technique referred to as transfer learning. This result will enable users to efficiently determine the insertion device settings for achieving arbitrary beam characteristics.
\end{abstract}

\maketitle


\section{\label{sec:intro}Introduction\protect\\}

For decades, synchrotron light source facilities have produced highly brilliant and tunable photon beams for experiments across many scientific disciplines, in particular through the use of insertion devices (IDs). At the Canadian Light Source, the Quantum Materials Spectroscopy Center (QMSC) beamline uses an elliptically polarizing undulator (EPU) type ID with a magnetic period of \SI{180}{\milli\meter} to produce soft x-rays with variable polarization in the energy range of \SI{15}{\electronvolt} to \SI{200}{\electronvolt}.

In materials science, having the ability to probe the orbital structure of electronic states with linear and circular dichroism measurements is critical to understanding the underlying physics in the system under study. Angle-resolved photoemission is one technique that can extract additional information from a sample by utilizing arbitrary polarization at low photon energies \cite{Day2019}. However, 100\% circular polarization is difficult to achieve due to the beamline optics altering the polarization of low energy photons as they propagate from the ID to the experiment end station \cite{wurtz2014, marcouille2007}. This introduces the requirement for arbitrary polarization of the light at the EPU, along with the corresponding requirement of knowing the EPU operating parameters that will deliver photons of a certain energy and polarization on demand.

A planar ID has its gap as one degree of freedom. In this case it is straightforward to build a one dimensional look-up table relating the energy of the radiated photon beam to the device gap, where the look-up table is typically generated from magnetic or beam based measurements. Operating an EPU in arbitrary polarization requires a multi dimensional lookup table to relate its parameters to the energy and polarization state of the photon beam. Moreover, the overall system may drift over months or years, for example due to changes in characteristics of the undulator or the beamline optics. Look-up tables built from measured data are limited to replacing their data one point in the ID's configuration space at a time, and hence the total time necessary to (re)measure data for multi dimensional look-up tables from either beam based or magnetic data becomes prohibitively large. The measurement time can be sidestepped by instead computing the undulator output polarization at any arbitrary point in configuration space from a model, for example using RADIA \cite{elleaume1997,chubar1998}. However, such calculations remain time consuming and the result is then limited by the accuracy of the model. The most attractive outcome is a fast-executing model that can be calibrated from a measured data set that is small compared to the size of a multi dimensional look-up table. In this article we propose that neural networks can be just such a model, providing rapid accurate predictions of the beam characteristics from a complex undulator.

\section{\label{sec:overview}Background\protect\\}

\subsection{\label{sec:overviewEPU}Elliptically Polarizing Undulators and Polarization\protect\\}

\begin{figure}[htp]
\includegraphics[width=1.0\textwidth]{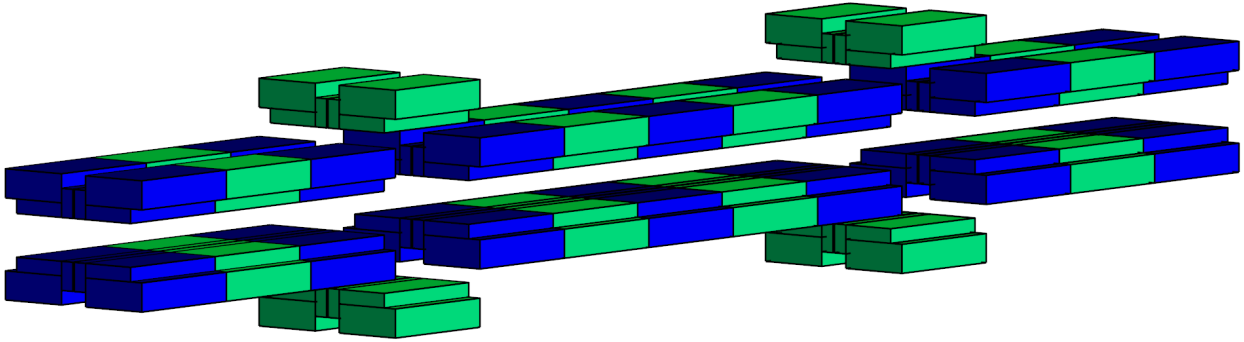}
\caption{\label{fig:quasiUndulator} Section of the QMSC insertion device (EPU \SI{180}{\milli\meter}) to illustrate magnet array.}
\end{figure}

The QMSC undulator is a quasiperiodic APPLE-II type EPU. A section of its modelled magnet arrays is shown in Figure \ref{fig:quasiUndulator}. Certain magnet blocks are offset vertically to incorporate a quasiperiodic magnetic structure, which reduces contamination of the harmonics present in the undulator spectrum \cite{chavanne2001}. Gap adjustments symmetrically change the vertical distance between the upper and lower magnet arrays. Independent longitudinal motion of the four girders are described with two independent parameters for operating the device, called the elliptical phase $\varphi_{E}$ and composite linear phase $\varphi_{L}$ \cite{sigrist2019}. By adjusting these three operating parameters, ($gap$, $\varphi_{E}$, $\varphi_{L}$), the strength and orientation of the undulator's magnetic field can be controlled, which in turn controls the energy and polarization of the radiated photons.

The polarization of the light radiated from the EPU can be described using the Stokes parameters S1, S2, and S3. For this application, the Stokes parameters are normalized and dimensionless, satisfying Equation \ref{stokesSum}, where each parameter ranges from \numrange{-1}{1}.

\begin{equation}\label{stokesSum}
S1^2 + S2^2 + S3^2 = S0^2 \leq 1 
\end{equation}

\subsection{\label{sec:overviewML}Machine Learning\protect\\}

Machine learning (ML) techniques have been studied for various particle accelerator applications. Recently, ML-based surrogate models have obtained accurate and fast-executing representations of the relevant beam dynamics from a sparse sampling of the physics simulation \cite{edelen2020}. Neural networks (NN), a sub-type of ML, have been trained to automatically tune and control large complex systems such as particle accelerators and insertion devices \cite{leeman2019,scheinker2019}. Their ability to be trained off-line using simulation data from computationally expensive codes and updated with measurement data has been demonstrated for multiple applications \cite{edelen2018, edelen2020}. This type of ML learning algorithm is referred to as supervised learning because the model is trained on labelled data sets. In this sense, ground truth outputs exist for each input \cite{arpaia2021}. In contrast to the simulation software from which ML models are trained, ML models can execute in fractions of a second with comparable accuracy in predicting the resulting beam parameters \cite{edelen2020}. Additionally, the ability of ML models to be updated with new measurement data ensures that they remain accurate as the characteristics of the modelled device changes \cite{edelen2020}.

With these advantages in mind, accomplishing the objective of this work entails acquiring a large training data set from simulations. ML models are able to learn complex nonlinear relationships using large amounts of training data, however, producing a large training data set is computationally expensive \cite{leeman2019}. In practice, the training data size depends on the complexity of the problem and complexity of the ML algorithm. Similar ML scenarios determined the amount of training data required by empirically evaluating the performance of their models with respect to the number of data points \cite{edelen2020}. This technique was used to determine the size of the required training data set. By varying the resolution of the ID settings in the training data, the size of the data set would change without affecting the equal representation of the operating modes of the ID within the data. The difficulty for ML models to interpolate between training points increases for complex, many parameter systems \cite{scheinker2019}, therefore the data size was chosen such that  the ID settings have sub-millimeter resolutions.

\section{\label{sec:methods}Methods\protect\\}

\subsection{\label{sec:methodPeriodic}Modelling the Undulator as Periodic Device\protect\\}

As an initial proof of concept for this work, training data was generated from a RADIA model of the undulator built as a periodic device. In this simplified case, the photon beam characteristics are derived from the undulator's effective and nominal fields \cite{sigrist2019} using Equations \ref{eqn:energy}, \ref{eqn:S1}, \ref{eqn:S2}, \ref{eqn:S3}  \cite{sigrist2018}. 

\begin{figure}[htp]
\includegraphics[width=1.00\textwidth]{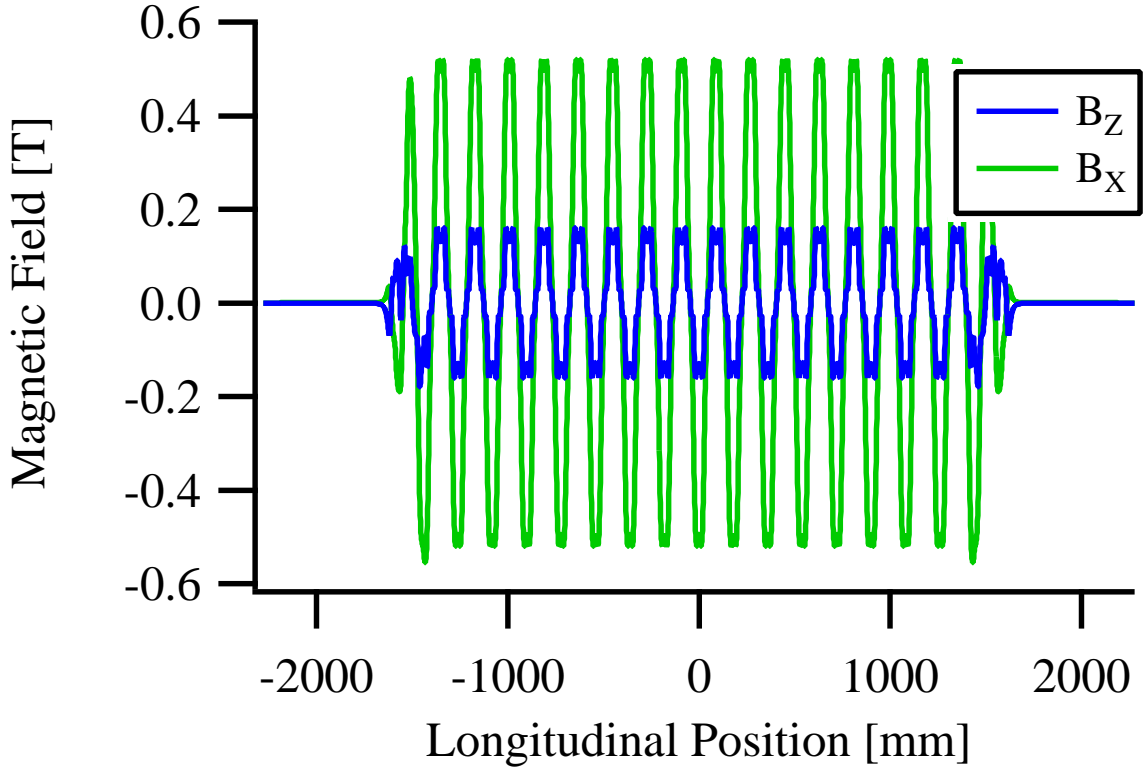}
\caption{\label{fig:fieldProfile} Magnetic field profile in EPU \SI{180}{\milli\meter}, modelled as a periodic device, calculated at $gap = \SI{15}{\milli\meter}$, $\phi_{E} = \SI{0}{\milli\meter}$, and $\phi_{L} = \SI{-60}{\milli\meter}$.}
\end{figure}

The effective field is an approximation of the undulator's peak field, $\hat{B}$, and is obtained via Fourier series decomposition of the modelled field profiles, $B_{x}(y)$ and $B_{z}(y)$. An example of one of these magnetic field profiles is shown in Figure \ref{fig:fieldProfile}. Equation \ref{eqn:effective} shows the Fourier series decomposition over harmonic i. The effective field is specific to an EPU's ($gap$, $\varphi_{E}$, $\varphi_{L}$) settings. Nominal fields are gap dependent phase independent terms, per Equation \ref{eqn:nominal}. $B_{z0}$ is given by $B_{zeff}$ at a horizontal polarisation and similarly $B_{x0}$ by $B_{xeff}$ at a vertical polarisation, such as $\varphi_{E} = \pm\lambda/2$. See Table \ref{tab:terms} for a list of the variables.

\begin{table}[htp]%
\caption{\label{tab:terms}%
Magnetic field terms and parameters \cite{sigrist2018}.
}
\begin{ruledtabular}
\begin{tabular}{lll}
Symbol & Term & Unit \\  
\colrule
$ E_{\gamma} $ & Photon Energy & \SI{}{\electronvolt} \\
$ E $ & Electron Energy & \SI{}{\giga\electronvolt} \\
$ \lambda $ & Undulator Period & \SI{}{\milli\meter} \\
$ k $ & Wave Number $ = 2\pi {\lambda}^{-1} $ & ${\SI{}{\milli\meter}}^{-1}$ \\
$ \phi_{E} $ & Elliptical Phase  & \SI{}{\milli\meter} \\
$ \phi_{L} $ & Linear Phase  & \SI{}{\milli\meter} \\
$ K $ & Deflection Parameter &  \\
$ {B}_{x,z eff} $ & Effective Horizontal, Vertical Field & \SI{}{\tesla} \\
$ {B}_{x,z0} $ & Nominal Horizontal, Vertical Field & \SI{}{\tesla} \\
\end{tabular}
\end{ruledtabular}
\end{table}

\begin{equation}\label{eqn:energy}
\begin{cases}
E_{\gamma} \big[ \SI{}{\electronvolt} \big] = 9.50 \frac{E^2 [\SI{}{\giga\electronvolt}]}{\lambda [\SI{}{\milli\meter}](1+ \frac{K^2}{2})}
\\
K = 0.0934 \lambda \big[ \SI{}{\milli\meter} \big] \hat{B} \big[ \SI{}{\tesla}\big]
\\
\hat{B} \approx B_{xeff}^2 + B_{zeff}^2

\end{cases}
\end{equation}

\begin{equation}\label{eqn:effective}
B_{x,zeff} = \sqrt{\sum_{x,zi}^{\infty} \frac{b_{i}^2}{i^2}}
\end{equation}

\begin{equation}\label{eqn:nominal}
B_{x,z0} = M e^{c_{1} \cdot \frac{gap^1}{\lambda^1} + c_{2} \cdot \frac{gap^2}{\lambda^2} + ...} 
\end{equation}

\begin{equation}\label{eqn:S1}
S1 = \frac{B_{xeff}^2 - B_{zeff}^2}{B_{xeff}^2 + B_{zeff}^2}
\end{equation}

\begin{equation}\label{eqn:S2}
S2 = 
\begin{cases}
\frac{1}{2} \frac{B_{x0} B_{z0} \sin^2 k\phi_{L}}{B_{xeff}^2 + B_{zeff}^2} & \text{if } \phi_{L} \geq 0
\\
-\frac{1}{2} \frac{B_{x0} B_{z0} \sin^2 k\phi_{L}}{B_{xeff}^2 + B_{zeff}^2} & \text{if } \phi_{L} < 0
\end{cases}
\end{equation}

\begin{equation}\label{eqn:S3}
S3 = 
\begin{cases}
 \frac{\sqrt{16{B}_{xeff}^2{B}_{zeff}^2 - {B}_{x0}^2{B}_{z0}^2 \sin^4 k\phi_{L}}}{ 2( {B}_{xeff}^2 + {B}_{zeff}^2 ) } & \text{if } \phi_{E} \geq 0
\\
 \frac{\sqrt{16{B}_{xeff}^2{B}_{zeff}^2 - {B}_{x0}^2{B}_{z0}^2 \sin^4 k\phi_{L}}}{ 2( {B}_{xeff}^2 + {B}_{zeff}^2 ) } & \text{if } \phi_{E} < 0
\end{cases}
\end{equation}

\begin{figure}[htpb]
\includegraphics[width=1.0\textwidth]{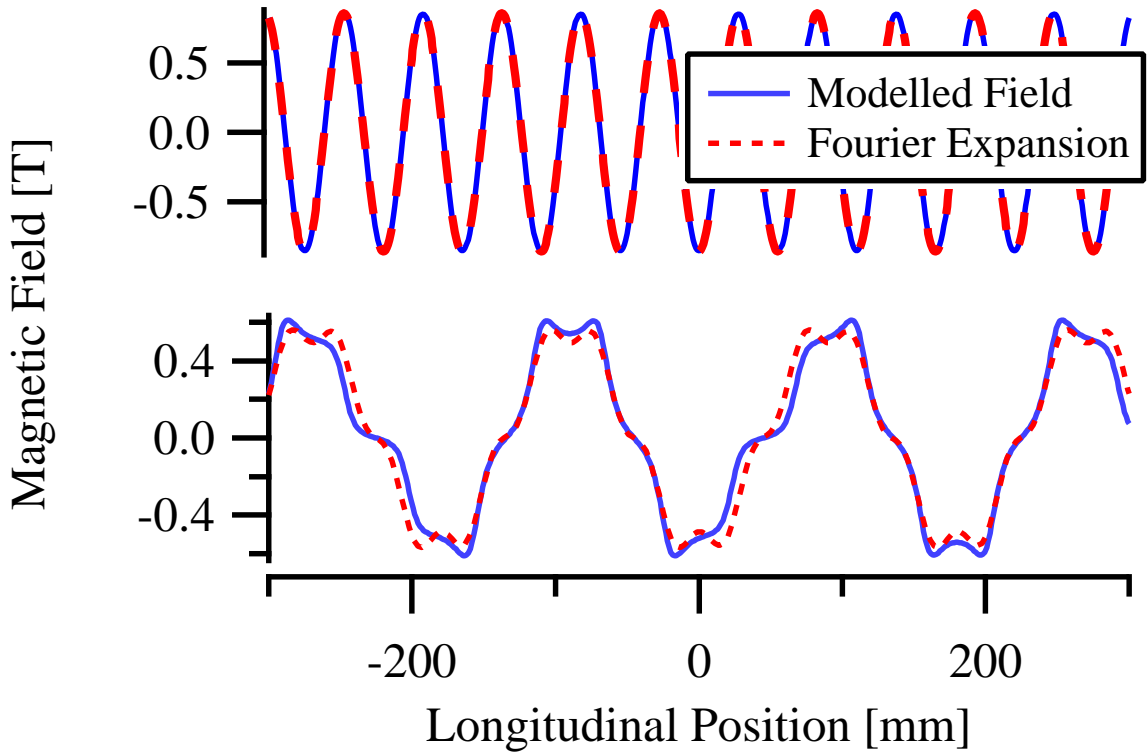}
\caption{\label{fig:fieldProfile2} Magnetic field profile (solid) and effective field (dashed) calculated from a Fourier expansion thereof. Top: \SI{55}{\milli\meter} periodic EPU. Bottom: \SI{180}{\milli\meter} quasi-periodic EPU.}
\end{figure}

However, describing the field profile in terms of effective fields introduces an approximation that holds poorly for quasiperiodic undulators. This point is illustrated in Figure \ref{fig:fieldProfile2}, which shows modelled undulator fields and their Fourier-determined effective equivalents for two cases. The upper plot shows a periodic undulator with a \SI{55}{\milli\meter} period, where the effective field closely matches the undulator field; the lower plot shows the \SI{180}{\milli\meter} quasiperiodic device under consideration, where the effective and undulator fields do not match. Calculating photon energy for the $n=1$ harmonic from the effective field for this configuration yields \SI{10.7}{\electronvolt}, whereas a more direct calculation (see next paragraph) yields \SI{9.4}{\electronvolt}. These results differ by 12\%, which highlights the inapplicability of Fourier decomposition for studying quasiperiodic fields.

\subsection{\label{sec:methodQPeriodic}Modelling the Undulator as Quasiperiodic Device\protect\\}

The photon beam characteristics can be determined without the approximation inherent to the effective field. This is achieved by modelling the undulator in its quasiperiodic configuration and exporting magnetic field data for analysis in the Synchrotron Radiation Workshop (SRW) code \cite{chubar1998_2}. Undulator radiation spectra are calculated at an observation window \SI{8}{\milli\meter} by \SI{8}{\milli\meter} in size and \SI{18}{\meter} downstream of the undulator; the calculation uses a filament electron beam without consideration of the actual beam emittance or energy spread. The spectra are calculated separately for the total (S0), horizontal (\SI{0}{\degree}), vertical (\SI{90}{\degree}), inclined linear (\SI{45}{\degree}, \SI{135}{\degree}), and left- and right- circular polarizations. Stokes parameters are then obtained by comparing the flux at the $n = 1$ harmonic for the different polarizations.

Scripting was developed in IGOR Pro to generate the large data set for training ML models \cite{wavemetrics}. The script can import and process magnetic field data for any number of EPU configurations. For each configuration, undulator radiation spectra are computed across an energy range near the undulator's $n = 1$ harmonic; the expected energy is calculated using Fourier-determined effective fields. The precise photon energy of the $n = 1$ undulator harmonic is determined by fitting a curve to the total photon flux. The photon beam characteristics and their corresponding ID settings describe a single case for the ML model.

\begin{figure}[htp]
\includegraphics[width=1.0\textwidth]{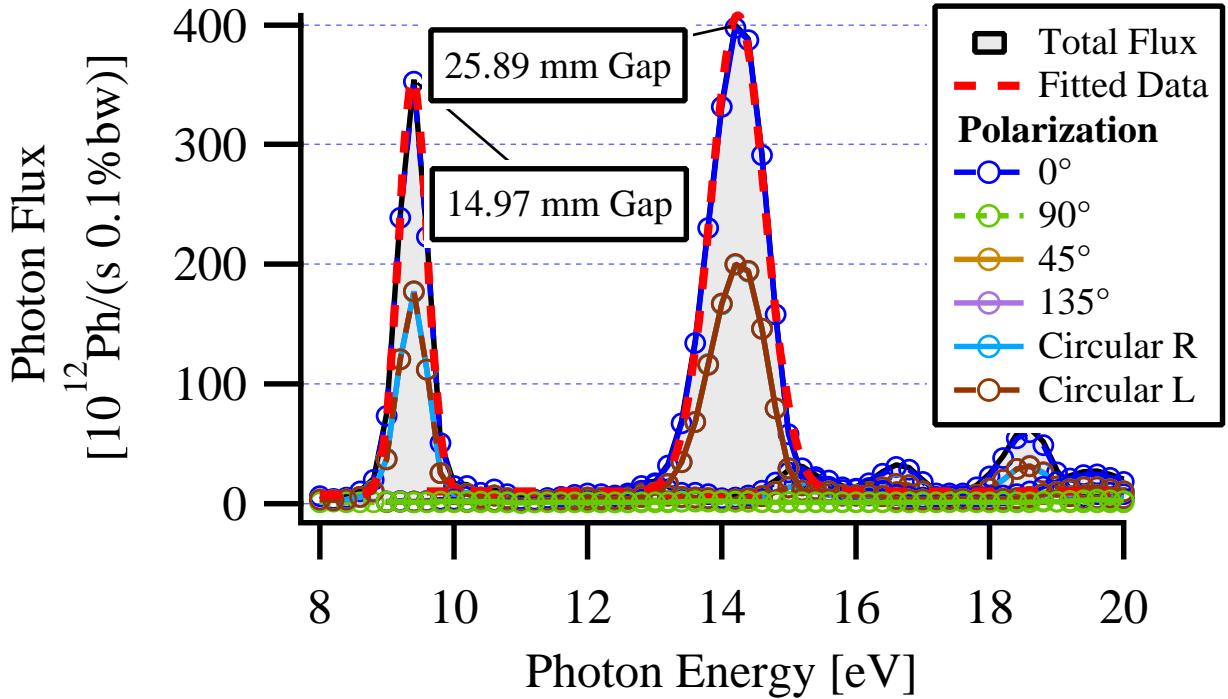}
\caption{\label{fig:radSpectra} Radiation spectra calculated for various polarization modes in SRW for two configurations of the quasiperiodic RADIA model. The configurations are at different gaps and are both in planar polarization ($\phi_{E} = \phi_{L} = \SI{0}{\milli\meter}$).}
\end{figure}

\begin{figure}[htp]
\includegraphics[width=1.0\textwidth]{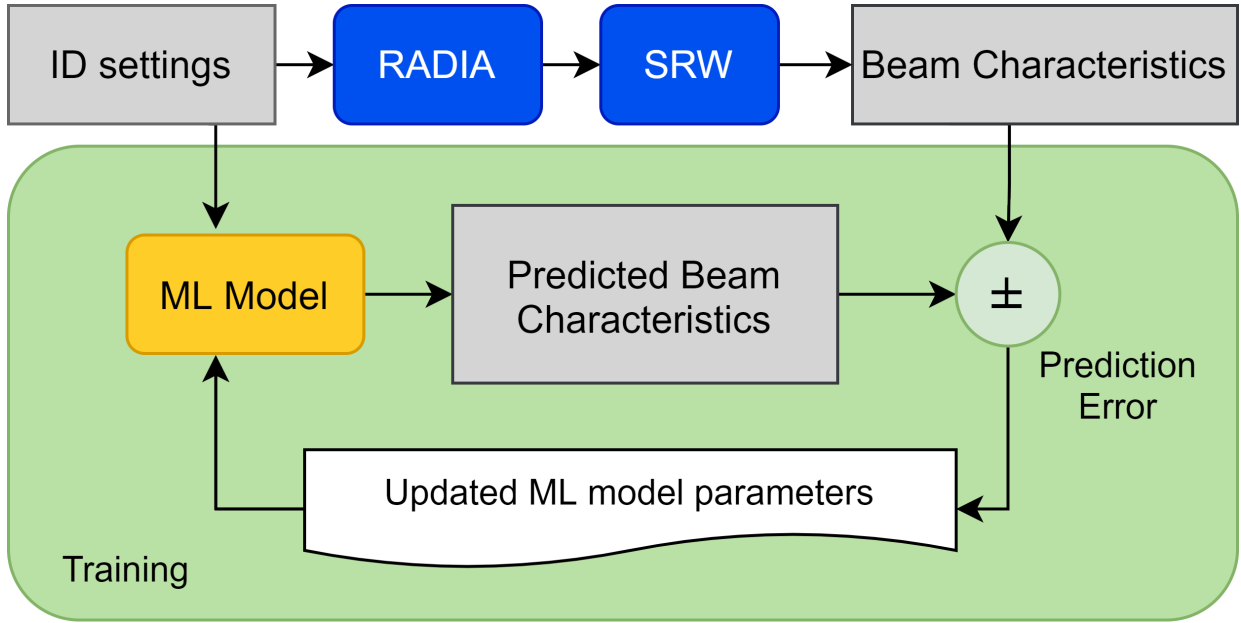}
\caption{\label{fig:InformationPipeline} The procedure for training a ML model using the physics simulation as the ground truth \cite{edelen2020}. The ML model parameters are dynamically updated during training. The hyperparameters of the model are manually adjusted until the model reaches a target performance.}
\end{figure}

Two example sets of undulator spectra are shown in Figure \ref{fig:radSpectra}. The script’s “information pipeline” and overall procedure for training a ML model is illustrated in Figure \ref{fig:InformationPipeline} \cite{edelen2020}.

Lastly, it is important to note that this methodology amounts to training a ML model based on the output of a RADIA model. Prior to this work, the RADIA model was refined with bench-based magnetic measurements of the actual undulator using a Hall probe and flipping coil setup. The RADIA model’s tuning process considered 45 EPU configurations, with priority given to planar and vertical polarization modes across various gap settings. Across the considered configurations, the typical relative difference between modelled magnetic fields compared to bench-based measurements on- and off-axis is 1\%.

\section{\label{sec:developNeural}Developing a Neural Network Model\protect\\}

A neural network model was created to predict photon beam energy as well as Stokes parameters S1, S2, S3; the model is hereafter referred to as NN4. The neural network was implemented using Keras with Tensorflow 2.0 backend and open-source scikit-learn packages \cite{abadi2016, pedregosa2011}.

A neural network is composed of individual neurons that accept multiple inputs and produce a single output. These neurons are arranged in layers to form a connected network \cite{smith1997}. The developed neural network is a feed-forward network in that the data propagates from input to output without looping between intermediate layers. The architecture of the neural network is a four hidden layer (128-64-32-16), fully connected neural network with a rectified linear unit activation function for each layer. The model was trained using backpropagation with the Adam optimizer \cite{lecun1989}. The mean squared error (MSE) is used both as a loss function and metric to monitor the performance of the model. The data set not used for training is divided equally, resulting in a 60-20-20 split of the training, validation, and testing data, respectively. The neural network model used scaled inputs in the range of (0, 1) and scaled outputs in the range (-1, +1).

Although a deep (many hidden layers) and wide (many nodes per layer) NN generally provides better fitting on training data, it is prone to overfitting \cite{leeman2019}. This issue was minimized by shuffling the data, implementing a learning schedule, adjusting the number of epochs (number of times the model is trained on a subset of data), and adjusting the batch size (the subset data size shown during training). 

\begin{figure}[htp]
\includegraphics[width=1.0\textwidth]{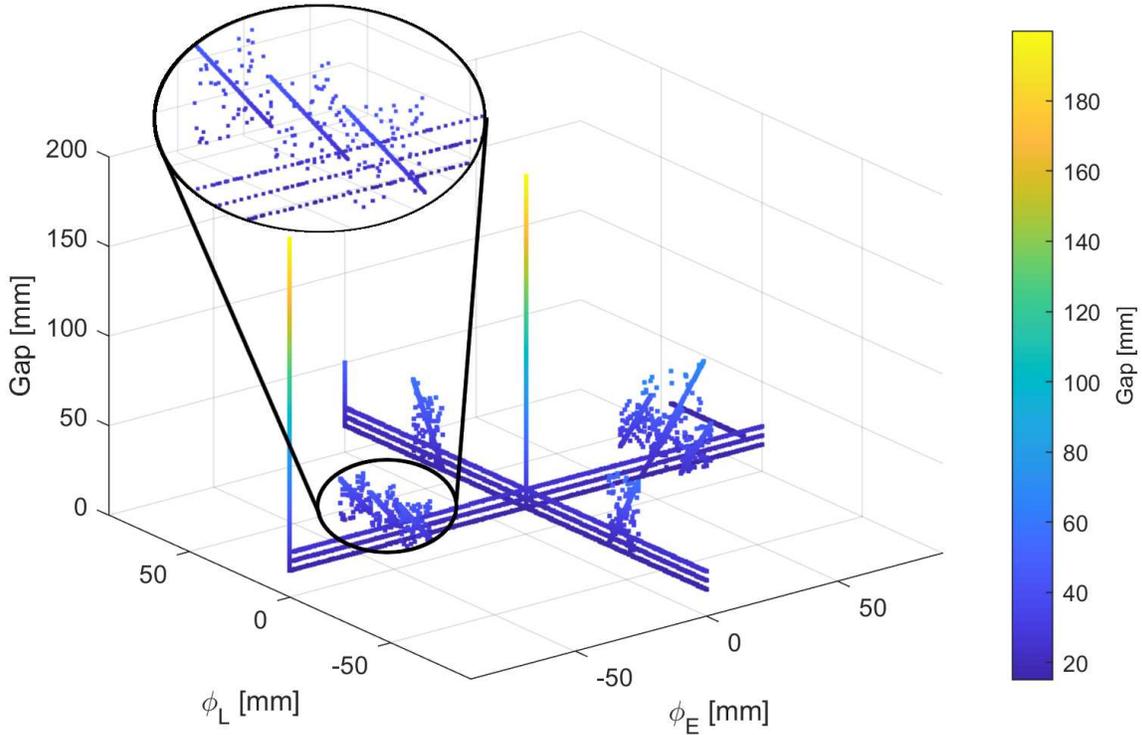}
\caption{\label{fig:graphCases} Distribution of 4175 simulated cases covering the operating modes of the EPU. A zoomed in portion of the figure is provided for clarifying that the apparent ‘lines’ on the figure are comprised of individual cases.}
\end{figure}

\begin{figure}[htp]
\includegraphics[width=1.0\textwidth]{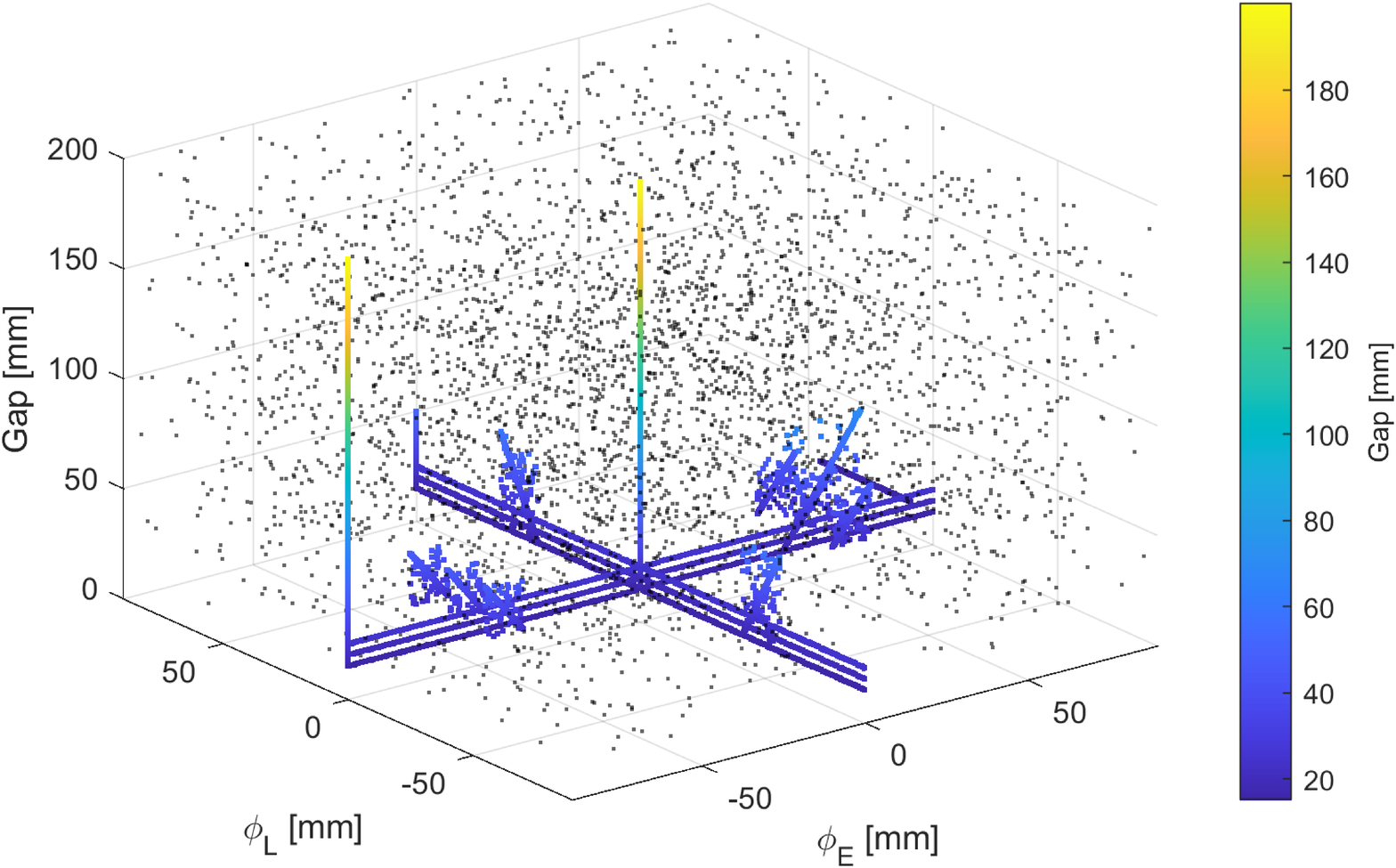}
\caption{\label{fig:totalCases} Total distribution of simulated data. The randomly generated cases are represented as black points to distinguish them from the initial 4175 cases.}
\end{figure}

The simulated training data for the NN model contained 4175 cases that sampled the EPU’s operating modes: planar, vertical ($\phi_{E}$ and $\phi_{L}$), circular (helicity 1 and 2), elliptical, linear, inclined (helicity 1 and 2), and a selection of Universal modes near circular at photon energies of interest to the beamline. To cover the total configuration space of the device, an additional 1000 cases were randomly generated for each quadrant formed by $\phi_{E}$ and $\phi_{L}$, for a total of 8175 cases. The 4175 and 8175 case data sets are shown in Figure \ref{fig:graphCases} and Figure \ref{fig:totalCases}, respectively. A single operating mode, E45:L45 ($\phi_{E} = \SI{45}{\milli\meter}$, $\phi_{L} = \SI{45}{\milli\meter}$, $gap = \SI{15}{\milli\meter}$), was not included in the smaller data set for reasons explained in Section \ref{sec:predicting}.

The train/test splitting technique for sorting the simulated data was employed with the ML models. This method entails dividing the data so that one group is used to train the model and a separate group is used to test the model. This computationally efficient approach was suitable because the data was shuffled prior to sorting, thereby guaranteeing the configuration space of the device was equally represented in the training and testing data sets. An appropriate distribution of the training and testing data is critical for effectively evaluating model performance. Note that this approach is equivalent to performing a k-fold cross-validation procedure with $k = 2$ \cite{stone1974}. The unscaled inputs occupy the following ranges: $gap =$ \SIrange{15}{200}{\milli\meter}; $\phi_{E} \text{ and } \phi_{L} = $ \SIrange{-90}{90}{\milli\meter}. The unscaled outputs occupy the following ranges: $E_{\gamma} = $ \SIrange{6}{400}{\electronvolt}, Stokes parameters \num{-1} to \num{1}. 

\section{\label{sec:predicting}Predicting Modelled EPU Beam Characteristics\protect\\}

The following results are drawn from the model’s performance on the simulated data sets for the EPU. The model used batch sizes of 16, a customized decaying learning rate schedule, and trained for 1500 epochs.

\subsection{\label{sec:configSpace}Configuration Space\protect\\}

\begin{figure}[htp]
\includegraphics[width=1.0\textwidth]{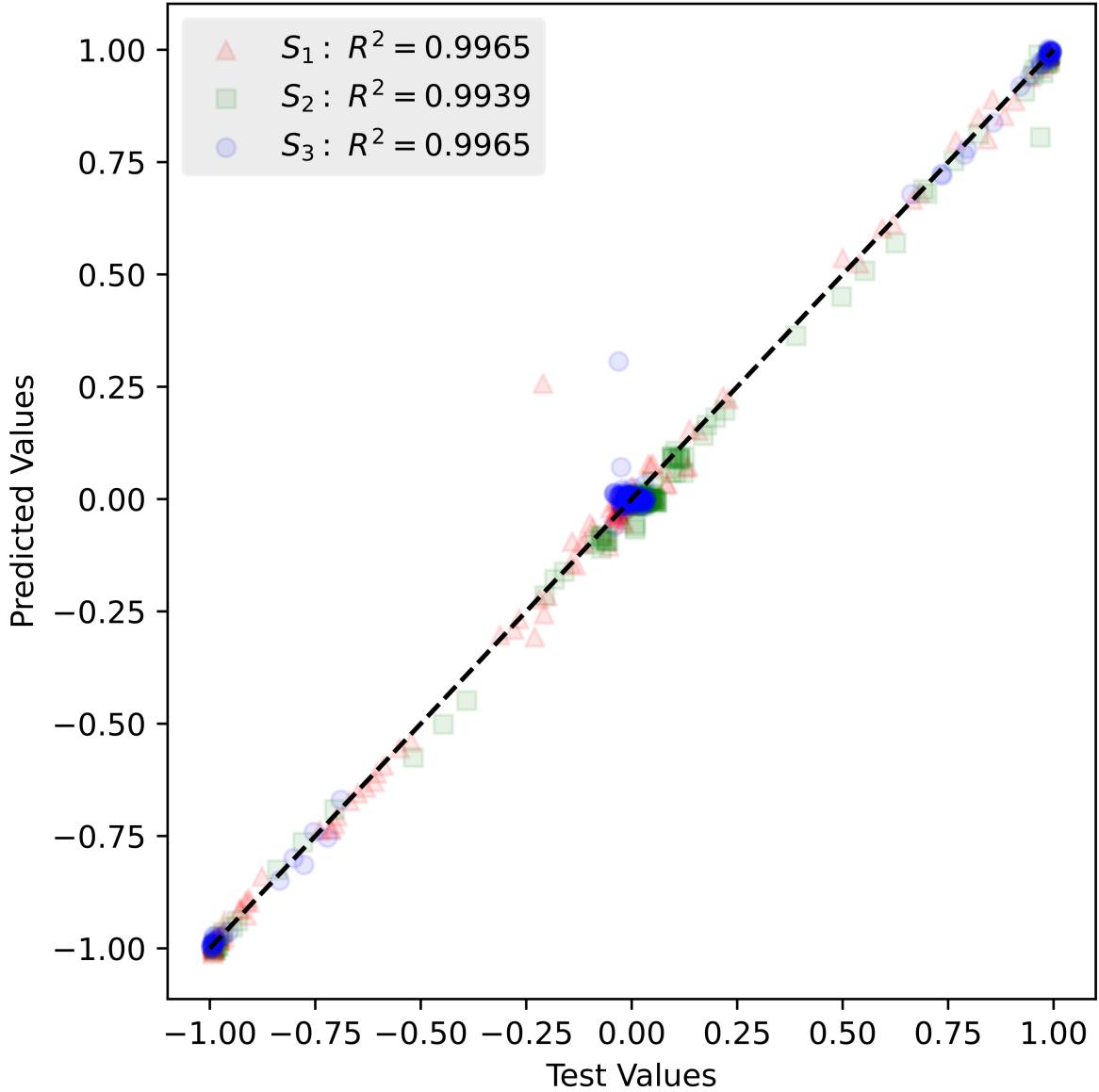}
\caption{\label{fig:e45L45_case2} Comparison between predicted and test output Stokes parameters for measured data from the NN. Excellent agreement shown for majority of test cases except for the E45:L45 case (predicted S1 value $\sim$ 0.26 versus a test value of $\sim$ -0.21, predicted S2 value of $\sim$ 0.8 versus a test value of $\sim$ 0.97, predicted S3 value $\sim$ 0.31 versus a test value of $\sim$ 0).}
\end{figure}

The first iteration of the NN model was trained on the simulated operating-modes data set, totalling 4175 unique cases. After testing the model on cases from the measured data set, it was apparent that the model did not generalize well to domains in configuration space not covered in the training data. Particularly, a single operating mode (E45:L45) contained in the measured data set, was not encompassed by the domain of the simulated operating-modes data set (Figure \ref{fig:graphCases}) used to train the model. This E45:L45 case was intentionally set aside from the simulated operating-mode data set to observe the model's ability to extrapolate for new domains. The predictions made by the NN model on the measured data set are shown in Figure \ref{fig:e45L45_case2} to demonstrate how the single E45:L45 case stands apart from other predicted cases. Although strict agreement between the predicted and test cases is not expected because the simulated and measured data sets are inherently unique, general agreement is expected. 

To test the prediction that the model requires training on each domain for which it will be tested, the model was trained on the operating-modes data set (4175 cases) with one additional quadrant of randomly generated data. The model accurately predicted the E45:L45 case when the extra quadrant data encompassed the E45:L45 case and poorly otherwise. 

Since we desire a ML model that may be used to predict the EPU beam characteristics for any operating mode, current or future, the second iteration of training the model was performed on the complete data set shown in Figure \ref{fig:totalCases}, which will be referred to as the simulated data set from now on. 

\subsection{\label{sec:results}Results\protect\\}

The ML model was evaluated based on its MSE, mean norm of the Stokes error vector (MSEV) shown in Equation \ref{eqn:msev}, the variance of their relative error in predictions, the mean absolute percentage error (MAPE) for the predicted photon energies and whether it satisfied the QMSC beamline’s error threshold shown in Equation \ref{eqn:stokesError}.

\begin{equation}\label{eqn:msev}
\text{MSEV} = \frac{1}{n} \sum_{i}^n || \vec{S}_{pred_i} - \vec{S}_{true_i} ||
\end{equation}

Equation \ref{eqn:msev} uses the Stokes parameters in vector notation where $\vec{S}_{pred}$  and $\vec{S}_{true}$ are the predicted and target Stokes vectors, respectively. The MSEV then represents the norm of the Stokes error vector averaged over $n$ test cases.  

\begin{equation}\label{eqn:stokesError}
\begin{cases}
\text{ if } |S3| = 1
\\
\text{ residual: } S1^2 + S2^2 \leq 0.03

\end{cases}
\end{equation}

Equation \ref{eqn:stokesError} states that for operation in circular mode, $|S3| = 1$, the residual components of the Stokes vector must be less than a threshold of 0.03. 

To evaluate how well the model generalized to new regions of the input configuration space, the predictions made by the model were compared with the target values. This step is performed using the testing data, which is ‘unseen’ by the model during its training. The NN model was compiled 30 times to establish its average performance. A summary of the model’s performance at predicting EPU beam characteristics from simulated data is given in Table \ref{tab:modelPerformance}.

\begin{table}[htp]%
\caption{\label{tab:modelPerformance}%
The model’s performance in predicting EPU beam characteristics. The metrics in this table are computed on the scaled predicted values which lie in the range (\num{-1} to 1). The metrics of the best performing model from the compilation are included to illustrate the model's performance potential. 
}
\begin{ruledtabular}
\begin{tabular}{lll}
\textrm{Model}&
\textrm{MSE}&
\textrm{MSEV}\cr\cr
\colrule
NN4 & Avg: $2.86 \times 10^{-4}$ & Avg: $2.01 \times 10^{-2}$ \\
& Best: $2.26 \times 10^{-4}$ & Best: $1.76 \times 10^{-2}$ \\
\end{tabular}
\end{ruledtabular}
\end{table}

\begin{figure}[htp]
\includegraphics[width=1.0\textwidth]{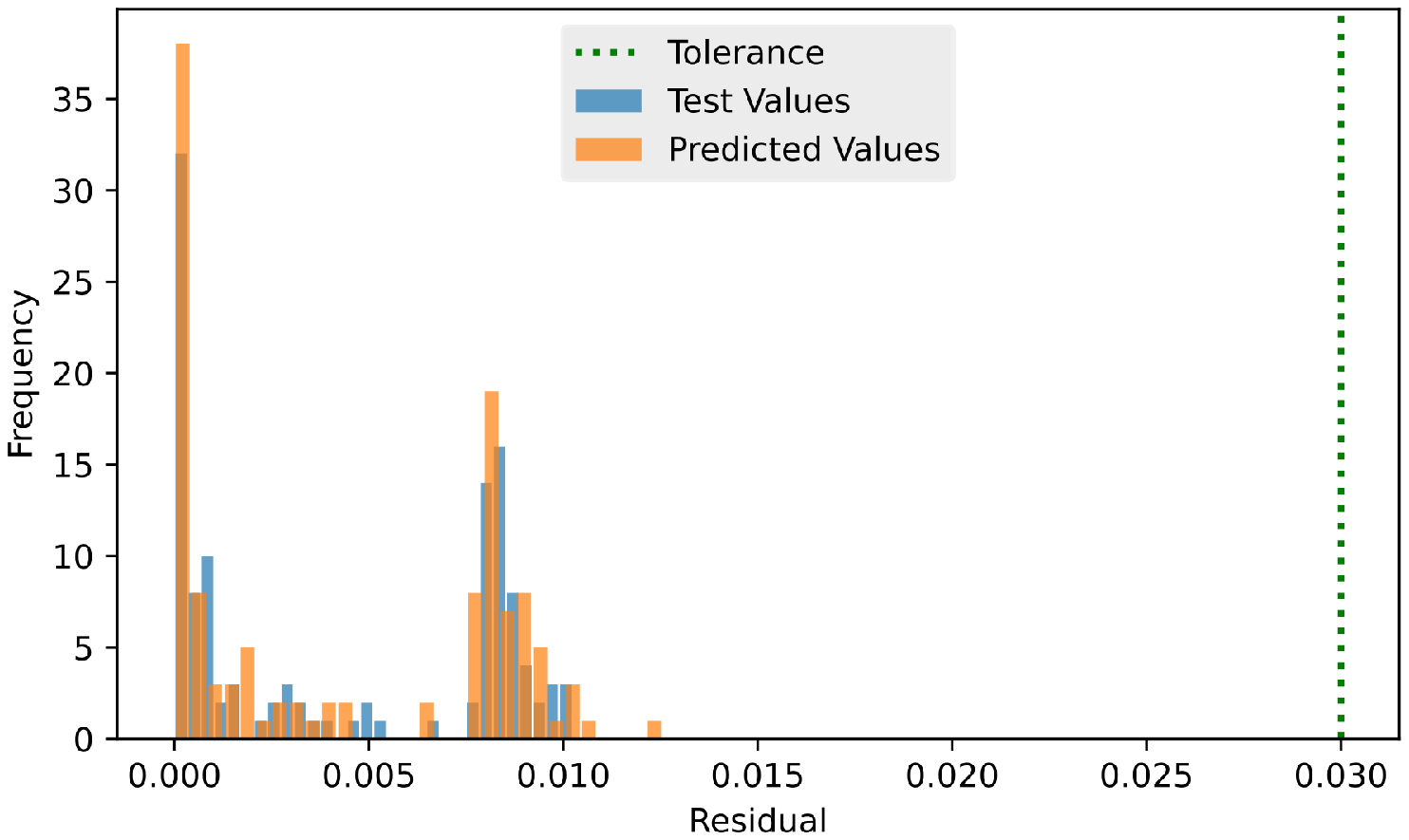}
\caption{\label{fig:s3Case} Histogram comparing the residuals for predictions on $|S3| = 1$ cases by the NN4 model with the residuals from the test data set. These cases correspond to cases where the test value of $|S3| = 1$ within a tolerance of $\pm 0.01$.}
\end{figure}

The QMSC beamline’s error threshold was satisfied by the model which is shown in Figure \ref{fig:s3Case}. The residuals from predictions follow the same trend (magnitude and frequency) as the residuals from the test values.

The accuracy of the 1633 predictions made by the NN4 model on the test data is shown in Table \ref{tab:NN4TargetProperties}. A regression score, R\textsuperscript{2}, is calculated for each output beam characteristic to indicate the correlation between predicted and test values. The variances of the relative errors from the predicted test cases of the NN4 model are included to represent the distribution of errors. The averaged MAPE indicated the NN4 model predicted the photon energy within 2.80\%.  The near-unity R\textsuperscript{2} values for each output indicate the NN4 model accurately predicts the EPU beam characteristics. The small variances of the relative errors in predictions indicate the predictions made by the NN4 model are tightly distributed around the mean (zero).

\begin{table}[htp]%
\caption{\label{tab:NN4TargetProperties}%
The properties of the model’s predicted beam characteristics. The regression scores are based on the scaled predicted values and the variance scores are computed from the un-scaled predicted values.}
\begin{ruledtabular}
\begin{tabular}{lll}
\textrm{Target}&
\textrm{R\textsuperscript{2}}&
\textrm{Variance $\sigma\textsuperscript{2}$ }\cr
\colrule
$E\gamma$ & $0.9998$ & $ 4.87 \:eV^{2} $ \\
$S1$ & $0.9992$ & $3.49 \times 10^{-4}$ \\
$S2$ & $0.9982$ & $2.68 \times 10^{-4}$ \\
$S3$ & $0.9986$ & $3.88 \times 10^{-4}$ \\
\end{tabular}
\end{ruledtabular}
\end{table}

\section{\label{sec:updating}Updating the Neural Network Model using Transfer Learning\protect\\}

The ability of a ML model to be updated with new data, as mentioned in Section \ref{sec:intro}, was investigated to determine if the model could predict the Stokes parameters derived from the magnet measurement data. The limited measured data mentioned in Section \ref{sec:methods} was used to update the ML model. 

Since the beam characteristics are only slightly different between the simulated and measured data sets, and the measured data set is small, the Calibrated neural network model used the entire NN4 model as the base model. This methodology involving the bottleneck layer of a trained model in transfer learning applications has been demonstrated by several computer vision works \cite{9127390}. Since the magnet measured data set is a small subset of the EPU configuration space, the updated model fits the measured data better with fewer trainable parameters. The measured data set contains 169 cases that proportionally represent the operating modes of the device. The Calibrated NN model was trained on 60\% of this data (101 cases) and tested on the remaining 40\% (68 cases). The model used batch sizes of 4, a customized decaying learning rate schedule, and trained for 300 epochs.

\begin{figure}[htp]
\includegraphics[width=1.0\textwidth]{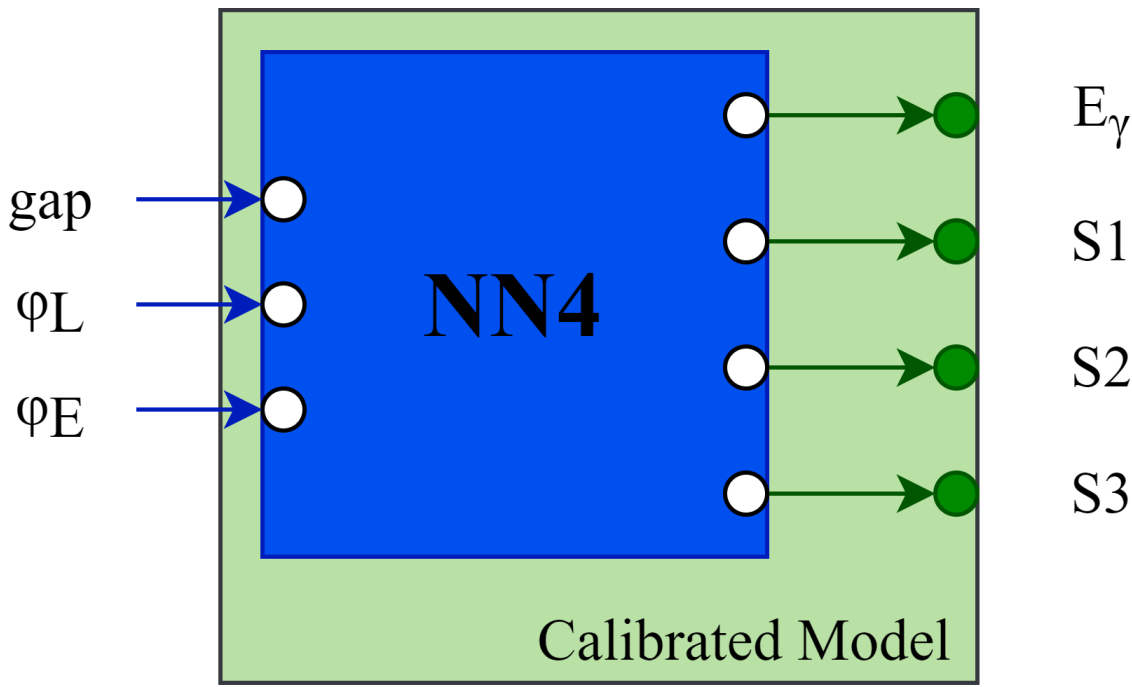}
\caption{\label{fig:architecture} Architecture of Calibrated NN model. The NN4 model acts as the base model with 4 outputs and the additional layer is added on top of NN4 (shown in green).}
\end{figure}

The updated neural network adds one additional layer, identical in structure to the  base model output layer (size 4, fully connected, using the linear activation function). A diagram is provided in Figure \ref{fig:architecture} to show the architecture of the Calibrated NN model. Similar techniques as those described in Section \ref{sec:developNeural} were employed to optimize the Calibrated NN model.

\subsection{\label{sec:transferResults}Transfer Learning Results\protect\\}
 
The updated neural network model was evaluated with the same metrics described in Section \ref{sec:predicting}. It is important to note that the performance of the updated neural network model may only be compared to the NN4 model when they are evaluated on similar data sets. The updated model was compiled 60 times to establish an average performance on the measured data set and the results are listed in Table \ref{tab:resultsCalibrated} below. For reference, a separate neural network model was created and trained solely on the magnetic measurement data. This Limited NN model was optimized to fit the dataset and the results from 60 compilations of the model are also included in Table \ref{tab:resultsCalibrated}. The performance of the NN4 model on the measured data set, averaged over 60 trials, is included to demonstrate the performance improvement of the Calibrated NN model.

\begin{table}[htp]%
\caption{\label{tab:resultsCalibrated}%
Summarized performance of the neural networks on the measured data set. The metrics in this table are computed on the scaled predicted values which lie in the range (\num{-1} to 1). 
}
\begin{ruledtabular}
\begin{tabular}{lll}
\textrm{Model}&
\textrm{MSE}&
\textrm{MSEV}\cr\cr
\colrule
Limited NN & Avg: $1.03 \times 10^{-2}$ & Avg: $1.21 \times 10^{-1}$ \\
& Best: $3.21 \times 10^{-3}$ & Best: $7.81 \times 10^{-2}$ \\
Calibrated NN & Avg: $1.31 \times 10^{-4}$ & Avg: $1.86 \times 10^{-2}$ \\
& Best: $9.72 \times 10^{-5}$ & Best: $1.57 \times 10^{-2}$ \\
NN4 & Avg: $5.32 \times 10^{-4}$ & Avg: $3.57 \times 10^{-2}$ \\
& Best: $4.20 \times 10^{-4}$ & Best: $3.21 \times 10^{-2}$ \\
\end{tabular}
\end{ruledtabular}
\end{table}

Like the NN4 model’s performance on the simulated data, the Calibrated NN model also satisfied the QMSC beamline’s error threshold. The largest residual for the six $|S3| = 1$ test cases was 0.005. The updated model's prediction accuracy on the measured data set is included in Table \ref{tab:CalibratedTargetProperties}. 

\begin{table}[htp]
\caption{\label{tab:CalibratedTargetProperties}%
The properties of the model’s predicted beam characteristics from the 68 test cases. The regression scores are based on the scaled predicted values and the variance scores are computed from the un-scaled predicted values}
\begin{ruledtabular}
\begin{tabular}{lll}
\textrm{Target}&
\textrm{R\textsuperscript{2}}&
\textrm{Variance $\sigma\textsuperscript{2}$ }\cr
\colrule
$E\gamma$ & $0.9995$ & $ 2.76 \:eV^{2} $ \\
$S1$ & $0.9997$ & $1.56 \times 10^{-4}$ \\
$S2$ & $0.9987$ & $1.96 \times 10^{-4}$ \\
$S3$ & $0.9997$ & $8.02 \times 10^{-5}$ \\
\end{tabular}
\end{ruledtabular}
\end{table}

\subsection{\label{sec:discussion}Discussion\protect\\}

The results in Table \ref{tab:resultsCalibrated} indicate how the Calibrated NN model outperforms the Limited NN model by roughly two orders of magnitude. This comparison demonstrates the necessity of applying transfer learning to a thoroughly trained base model. The improvement of the Calibrated NN model upon the NN4 model is identified by the smaller errors in predictions. However, the MAPEs for the Calibrated NN and NN4 models were 4.57 \% and 3.56 \%, respectively. This lack of improvement for the Calibrated NN model is unexpected because its smaller MSE indicates a better fit to the data than the NN4 model.

 The similar performance characteristics between Tables \ref{tab:NN4TargetProperties} and \ref{tab:CalibratedTargetProperties} indicate the Calibrated NN model has a comparable prediction accuracy on the measured data to the NN4 model on the simulated data. The near-unity regression scores in Table \ref{tab:CalibratedTargetProperties} indicate the model is accurately predicting the beam characteristics and the small variances imply the relative errors are small and closely distributed around zero.

\section{\label{sec:conclusion}Conclusion\protect\\}

The results of this work demonstrate the feasibility of generating a ML model to accurately predict the photon beam characteristics of a quasiperiodic EPU. More specifically, this work demonstrates the ability of a neural network to accurately model the complex, multi-parameter functions of an ID. This outcome was achieved by optimizing the neural network model to fit a large simulated data set. The importance of properly sampling the configuration space in the development of a neural network was also demonstrated by the E45:L45 case. 

Secondly, the successful application of transfer learning demonstrates how the neural network model was easily adapted to a measured data set. This stage was accomplished by building a separate neural network model, referred to as the Calibrated NN model, based off the NN4 model. This model was then trained on the limited magnetic measured data set to provide more accurate predictions of the radiated light at the ID.  The predictions produced by the Calibrated NN model satisfy the QMSC beamline’s error threshold and the relative errors in predictions were shown to be within an acceptable threshold. The photon energy was predicted more accurately by the NN4 than the Calibrated NN model, indicated by the slightly smaller averaged MAPE. However, this accuracy difference is small. The Calibrated NN model showed promising improvement in predicting the Stokes vector. The MSEV was determined to be $1.86 \times 10^{-2}$, indicating the predicted Stokes vectors closely agree with the test Stokes parameters.   

Thirdly, the deployment of this updated neural network model provides a synchrotron beamline with a fast-executing model for producing look up tables and/or predicting single ID cases.

Future work for this project includes the development of a ML model that will predict the beam characteristics at the end station by following a similar training and calibration approach. Polarization measurements will be acquired using a polarimeter located at the end station and used to calibrate a neural network from this work. The completion of this work will provide users with an efficient tool for predicting the end station beam characteristics for arbitrary ID configurations.

\section{\label{sec:ack}Acknowledgements\protect\\}

Research at the Canadian Light Source was funded by the Canada Foundation for Innovation, the Natural Sciences and Engineering Research Council of Canada, the National Research Council Canada, the Canadian Institutes of Health Research, the Government of Saskatchewan, Western Economic Diversification Canada, and the University of Saskatchewan. This research was enabled in part by support provided by WestGrid (www.westgrid.ca) and Compute Canada (www.computecanada.ca).

\bibliography{apssamp}

\end{document}